\documentclass[letterpaper, 12pt]{article}

\usepackage[hmargin=3.17cm,vmargin=2.54cm]{geometry}

\usepackage{graphicx}
\usepackage{amsfonts}
\usepackage{amssymb}
\usepackage{amsmath}
\usepackage{booktabs}
\usepackage{doi}
\usepackage{setspace}

\DeclareMathOperator{\diag}{diag}
\DeclareMathOperator{\minimize}{minimize}
\DeclareMathOperator{\subjectto}{subject\;to}

\begin{document}


\title{Inverse design and implementation of a wavelength demultiplexing grating coupler}
\author{Alexander Y. Piggott$^1$, Jesse Lu$^1$, Thomas M. Babinec$^1$, \\ Konstantinos G. Lagoudakis$^1$, Jan Petykiewicz$^1$, and Jelena Vu\v{c}kovi\'{c}$^{1\ast}$}
\date{}
\maketitle

\begin{center}
\vspace{-4ex}
$^1$ Ginzton Laboratory, Stanford University, Stanford, CA, 94305

$^\ast$ \emph{jela@stanford.edu}
\end{center}

\begin{abstract}
Nanophotonics has emerged as a powerful tool for manipulating light on chips. Almost all of today's devices, however, have been designed using slow and ineffective brute-force search methods, leading in many cases to limited device performance. In this article, we provide a complete demonstration of our recently proposed inverse design technique, wherein the user specifies design constraints in the form of target fields rather than a dielectric constant profile, and in particular we use this method to demonstrate a new demultiplexing grating.  The novel grating, which has not been developed using conventional techniques, accepts a vertical-incident Gaussian beam from a free-space and separates O-band $(1300~\mathrm{nm})$ and C-band $(1550~\mathrm{nm})$ light into separate waveguides. This inverse design concept is simple and extendable to a broad class of highly compact devices including frequency filters, mode converters, and spatial mode multiplexers.
\end{abstract}

Conventional integrated photonic devices \cite{jsun_nat2013} include components such as waveguide directional couplers \cite{pdtrinh_el1995}, multimode interference couplers \cite{lbsoldano_jlt1995}, distributed Bragg reflectors \cite{temurphy_jlt2001}, micro-ring resonators \cite{pdumon_iptl2004}, adiabatic tapers \cite{yshani_iqe1991} and grating couplers \cite{dtaillaert_jjap2006}. In all of these cases, the design space spans a relatively small $\left(\sim 2 - 5\right)$ number of parameters such as structure widths, heights and periodicity that are tuned throughout the photonic device design stage. To design a device, the photonic engineer specifies a dielectric profile, computes the electromagnetic field response using Maxwell’s equations, and compares the response to the device specifications. The process is then repeated, modifying the dielectric profile each iteration, until satisfactory performance is obtained. This brute force approach suffers from a long device design cycle, and does not take full advantage of the available design space of fabricable devices.

To this end, a wide variety of increasingly sophisticated approaches have been developed to search through this parameter space and optimize specific nanophotonic structures. Several methods, namely genetic algorithms \cite{agondarenko_oe2008, ahakansson_oe2005,mminkov_sr2014} and particle swarm optimization \cite{yma_oe2013}, ignore the underlying physics but have achieved considerable success in fine-tuning existing structures, and designing photonic crystal devices with selectively removed holes and posts. These methods, however, are typically restricted to optimizing a relatively small number of geometric parameters, and scale poorly with additional degrees of freedom. Other methods exploit the underlying physics to quickly converge on local optima, typically by computing the local gradient of a performance metric and using steepest-descent optimization \cite{sboyd_2004}. Owing to their much faster convergence, they can be used to design more complex structures with arbitrary topologies \cite{jsjensen_apl2004,piborel_oe2004,amutapcica_eo2009,jjensen_lpr2011,lalau-keraly_oe2013,aniederberger_oe2014,jlu_oe2013}. These algorithms can search through the design space of complex, aperiodic structures beyond those we can come up with based on our intuition and experience. Such devices may be able to provide novel functionality, or higher performance and smaller footprints than traditional devices, due to the greatly expanded design space.

We have recently proposed an inverse design approach for linear optical components, where the user specified input is not the basic structure of the device, but rather a set of performance metrics \cite{jlu_oe2013}. These include the device area, the modes of input and output waveguides to use (e.g. TE, TM, and mode order), extinction ratios (e.g. $> 20~\mathrm{dB}$), and insertion loss (e.g. $< 1~\mathrm{dB}$). More rigorously, the user only specifies the coupling efficiencies between a set of input and output modes at various frequencies. Any linear optical device can be specified in this fashion \cite{dabmiller_oe2012}, including mode converters, spatial mode multiplexers, and wavelength demultiplexers.

Two of the most important functions in integrated photonics, for both chip-to-chip and intra-chip optical interconnects, are wavelength-division multiplexing (WDM) and vertical-incidence coupling \cite{gtreed_2008}. A device combining these two functions in a single, compact layout could be particularly useful for coupling between silicon photonic layers in a stacked-die microprocessor, or for coupling on/off chip using optical fibers.  Although a uniform grating coupler with a tilted incident beam will act as a wavelength-demultiplexing grating coupler for two wavelength bands \cite{groelkens_oe2007}, the symmetry imposed by a vertically incident beam implies that a uniform grating coupler cannot split wavelengths. Indeed, an efficient vertically-incident wavelength-demultiplexing grating coupler cannot be designed using current analytic methods, or by tuning a small number of parameters by hand.

Here, we provide a full demonstration of the inverse design technique's power by experimentally demonstrating a vertical-incidence, wavelength-demultiplexing grating coupler fabricated in silicon-on-insulator (SOI). The grating accepts a vertically incident beam from free space or an optical fiber, and splits O-band $\left(\sim 1300~\mathrm{nm}\right)$ and C-band $\left(\sim 1550~\mathrm{nm}\right)$ light into separate silicon photonic waveguides with high extinction ratios $(> 10~\mathrm{dB})$.

\section*{Results}

In general, we can specify the performance of a linear optical device by defining the mode conversion efficiency between a set of input modes and output modes \cite{dabmiller_oe2012}. These modes are specified by the user, and kept fixed during the optimization process. The input modes $i = 1 \ldots M$ are at frequencies $\omega_i$, and can be represented by equivalent current density distributions $\mathbf{J}_i$. The generated electric fields $\mathbf{E}_i$ should satisfy Maxwell's equations in the frequency domain,
\begin{align}
\nabla \times \mu_0^{-1} \nabla \times \mathbf{E}_i - \omega^2 \, \epsilon \, \mathbf{E}_i = - i \omega_i \mathbf{J}_i, \label{eqn:fsg_physics_spec_1}
\end{align}
where $\epsilon$ is the electric permittivity, and $\mu_0$ is the magnetic permeability of free space.

We can then specify $N_i$ output modes of interest for each input mode $i$. The output mode electric fields $\mathcal{E}_{ij}$ are given over output surfaces $S_{ij}$, where $j = 1 \ldots N_i$. The amplitude of each output mode should be bounded between $\alpha_{ij}$ and $\beta_{ij}$, which can be expressed as
\begin{align}
\alpha_{ij} \leq \left| \iint_{S_{ij}} \mathcal{E}^\dagger_{ij} \cdot \mathbf{E}_i \mathrm{d}S \right| \leq \beta_{ij} \label{eqn:fsg_physics_spec_2}
\end{align}
for $i = 1 \ldots M$ and $j = 1 \ldots N_i$. We are thus interested in finding $\epsilon$ and $\mathbf{E}_i$ which simultaneously satisfy (\ref{eqn:fsg_physics_spec_1}) and (\ref{eqn:fsg_physics_spec_2}).

\begin{figure}
	\center
	\includegraphics[scale=0.45]{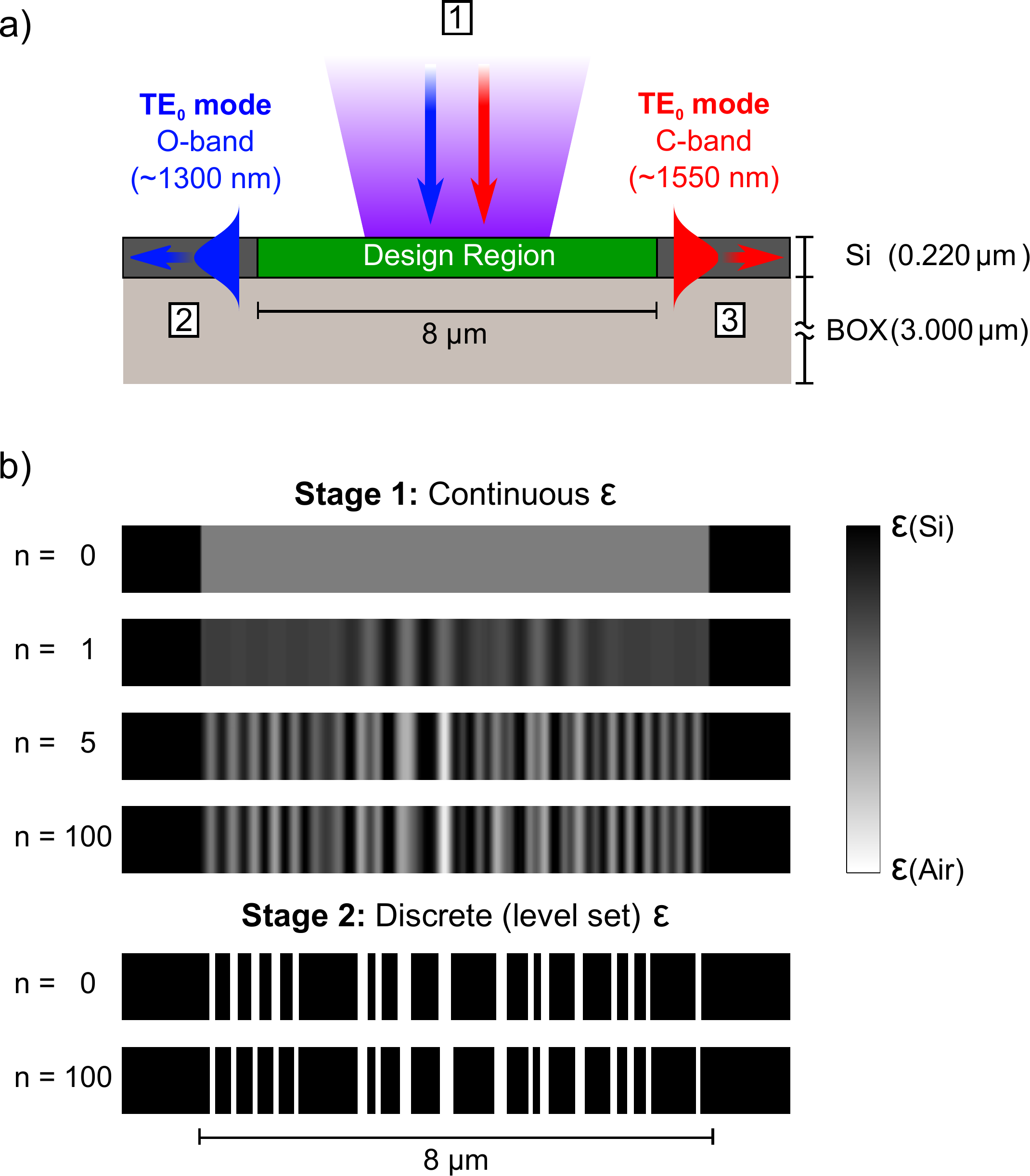}
	\caption{The inverse design procedure. \textbf{(a)} The device specifications provided to the inverse design algorithm, which consist of the input and output modes, the design region, and the surrounding structure. The device is fabricated by fully etching a $220~\mathrm{nm}$ silicon layer on $3~\mathrm{\mu m}$ of buried oxide, in the pattern produced by the optimization algorithm (see figure \textbf{1b}). \textbf{(b)} Intermediate steps in the optimization process, where $n$ is the iteration number. The optimization process proceeds in two stages. In the first stage, the permittivity $\epsilon$ is allowed to vary continuously. In the second stage, the design is converted to a binary level-set representation and fine-tuned. To clearly illustrate the design process, the diagrams are not to scale in the vertical direction. \label{fig:1_fsg_invdes}}
\end{figure}

The WDM grating was designed by specifying the input mode to be a $4.4~\mathrm{\mu m}$ diameter vertically-incident Gaussian beam, and the output modes to be the fundamental TE mode of the output silicon slab waveguides, as shown in figure \textbf{\ref{fig:1_fsg_invdes}a}. The algorithm was directed to maximize power into the left waveguide and minimize power into the right waveguide at $1300~\mathrm{nm}$, and the converse at $1550~\mathrm{nm}$. 

Given the design specifications, our algorithm iteratively optimizes the structure using a close analogue of steepest-descent optimization to meet the constraints given in equation (\ref{eqn:fsg_physics_spec_2}), as detailed in the Supplementary Information. We use finite-difference frequency domain (FDFD) simulations to calculate the local gradient during each step \cite{wshin_jcp2012, wshin_maxwell_webpage}. As illustrated in figure \textbf{\ref{fig:1_fsg_invdes}b}, the design procedure consists of two stages. In the first stage, the permittivity $\epsilon$ is allowed to smoothly vary within the design region. In the second stage, we convert the structure to a level-set representation \cite{sosher_2003} and fine-tune the final structure. We did not apply any design rules such as a minimum feature size, although such constraints could be incorporated into the design process \cite{jlu_oe2013}. The entire inverse design process for this grating took only $\sim 15$ minutes on a single Intel Core i7 processor.

The final device was fabricated in SOI with a $220~\mathrm{nm}$ thick Si device layer and a $3~\mathrm{\mu m}$ BOX (buried oxide) layer by fully etching the Si device layer, as described in figure \textbf{\ref{fig:2_fsg_struct}a}. The dimensions and locations of the grating trenches are detailed in Supplementary Table \textbf{1}. The broadband performance of the device was verified by using 2D finite-difference time domain (FDTD) simulations. The phasor fields at $1293~\mathrm{nm}$ and $1540~\mathrm{nm}$ for a $4.4~\mathrm{um}$ diameter Gaussian beam, obtained using 2D-FDTD simulations, are plotted in figure \textbf{\ref{fig:2_fsg_struct}b}. The fabricated device is presented in figure \textbf{\ref{fig:2_fsg_struct}c}. 

\begin{figure}
	\center
	\makebox[\textwidth][c]{\includegraphics[scale=0.45]{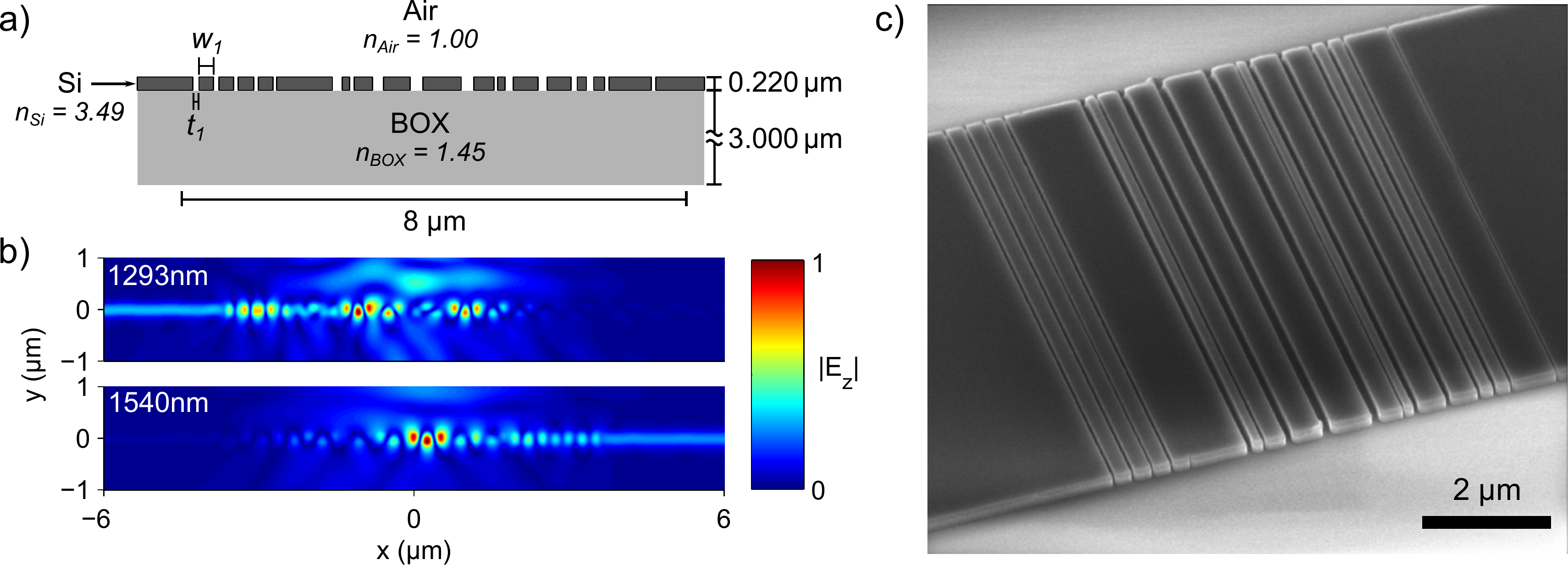}}
	\caption{Basic structure of the wavelength demultiplexing grating. \textbf{(a)} The device consists of a $220~\mathrm{nm}$ silicon layer with fully etched trenches on top of $3~\mathrm{\mu m}$ of buried oxide. Dimensions of the trench widths $\left( t_n \right)$ and spacings $\left( w_n \right)$, chosen by the optimization algorithm, are listed in Supplementary Table 1. \textbf{(b)} Frequency-domain electric field amplitudes for a $4.4~\mathrm{\mu m}$ diameter Gaussian beam incident on the center of the grating. At $1293~\mathrm{nm}$, light is only coupled into the fundamental mode of the left waveguide, whereas at $1540~\mathrm{nm}$, light is only coupled into the right waveguide. The fields plotted here were calculated using a finite-difference time-domain simulation, and post-processed using narrowband frequency filters to obtain the phasor fields. \textbf{(c)} Scanning electron microscopy (SEM) image of the fabricated device. \label{fig:2_fsg_struct}}
\end{figure}

\begin{figure}
	\center
	\makebox[\textwidth][c]{\includegraphics[scale=0.45]{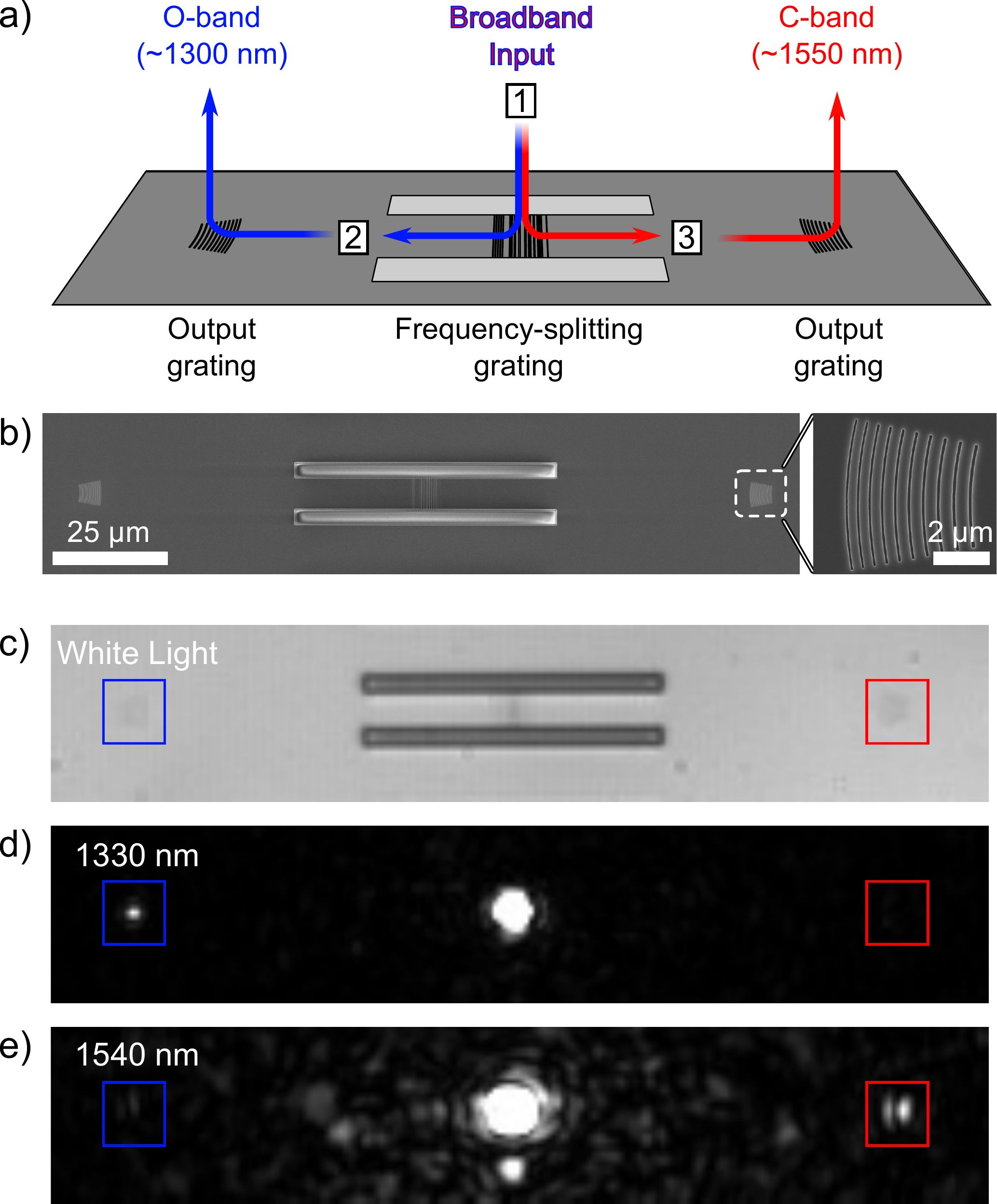}}
	\caption{Experimental configuration used to measure the frequency-splitting grating coupler. \textbf{(a)} The frequency-splitting grating is excited by a free-space beam, which couples light into the silicon slab waveguide. The coupled light is then scattered upwards by the two identical output gratings, and collected via free-space optics.
	\textbf{(b)} Scanning electron microscopy (SEM) images of the fabricated structures. The wavelength-demultiplexing grating coupler is in the center of a $8~\mathrm{\mu m}$ wide, $70~\mathrm{\mu m}$ long waveguide, and the two output grating couplers are placed $50~\mathrm{\mu m}$ from the ends of the waveguide. The output grating couplers are strongly chirped to provide broadband coupling, and slightly curved to minimize back-reflections into the waveguide.
	\mbox{\textbf{(c - e)}}~Infrared camera images of the device under \textbf{(c)} broad white-light illumination, and when the frequency-splitting is excited by a focused laser beam tuned to \textbf{(d)} $1320~\mathrm{nm}$ and \textbf{(e)} $1540~\mathrm{nm}$. The locations of the two output gratings are indicated by the colored boxes. Light is only coupled to the left output grating at $1320~\mathrm{nm}$, and only the right grating at $1540~\mathrm{nm}$. Back-scatter from the rear surface of the wafer is visible near the center of both images. \label{fig:3_fsg_expconfig}
}
\end{figure}

The final fabricated structures also incorporate a waveguide and curved broadband output grating on either side of the wavelength-demultiplexing grating, as shown in figure \textbf{\ref{fig:3_fsg_expconfig}a}. Scanning electron microscopy (SEM) images of the fabricated structures are shown in figure \textbf{\ref{fig:3_fsg_expconfig}b}. The output gratings were strongly chirped to provide broadband performance. To minimize Fabry-Perot resonances due to back-reflections, the output gratings were slightly curved and placed far from the ends of the waveguides (inset). In future, the output waveguides could be edge-coupled to optical fibers to obtain a well-characterized out-coupling efficiency \cite{ckopp_ijqe2011}.

The frequency-splitting grating was excited in the vertical direction by a focused Gaussian beam with a diameter of $\sim 4\mu m$, and a fraction of the coupled light was out-coupled by the output gratings. To ensure a clean Gaussian input beam, the source was passed through a length of single-mode optical fiber. Spectroscopic data was measured using a broadband LED source, and narrowband images were taken using tunable lasers. The structure was both excited and measured through a single plan-apochromat microscope objective integrated into a custom confocal microscopy setup. The collected light was either directly imaged using an InGaAs near-infrared camera, or spatially filtered by a pinhole at a focal plane and analyzed using a grating spectrometer with an InGaAs detector.

Images of the device broadly illuminated with white light, and excited by a focused tunable laser are presented in figure \textbf{\ref{fig:3_fsg_expconfig}c-e}.  At $1320~\mathrm{nm}$, light is only coupled to the left output grating, and at $1540~\mathrm{nm}$, light is only coupled to the right output grating, clearly demonstrating the basic functionality of the device.

\begin{figure}
	\center
	\includegraphics[scale=0.6]{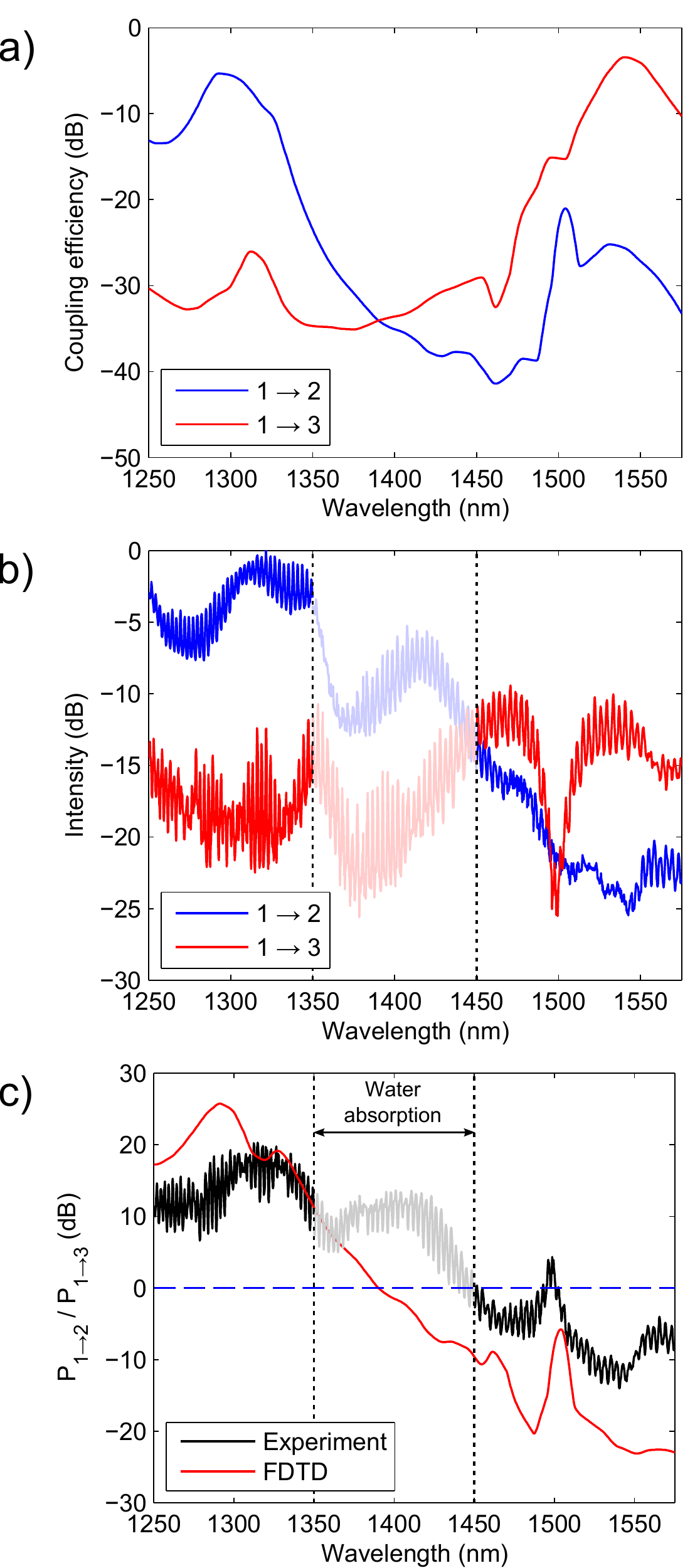}
	\caption{Simulated and measured coupling efficiencies of the frequency-splitting grating coupler. The device ports are labelled as in figure \textbf{\ref{fig:3_fsg_expconfig}a}. \textbf{(a)} Simulated coupling efficiency into the left and right waveguides, calculated using an finite-difference time domain (FDTD) simulation on the structure with optimized parameters. A $4.4~\mathrm{\mu m}$ diameter Gaussian beam was used as the input. \textbf{(b)} Experimental measured intensities from the left and right output gratings. The intensities are only normalized with respect to the source since the output grating efficiencies are not known.  Around $1300~\mathrm{nm}$, light is predominantly coupled to the left grating, whereas around $1550~\mathrm{nm}$, light is predominantly coupled into the right grating. Measurements in the $1350 - 1450~\mathrm{nm}$ band (shaded) are corrupted by water absorption lines in the atmosphere. \textbf{(c)} Simulated and experimentally measured splitting ratios for the coupler, defined as the the ratio of power into the left and right waveguides. We have experimentally measured splitting ratios of $17\pm2~\mathrm{dB}$ at $1310~\mathrm{nm}$ and $12\pm2~\mathrm{dB}$ at $1540~\mathrm{nm}$. \label{fig:5_fsg_expdata}}
\end{figure}

In figure \textbf{\ref{fig:5_fsg_expdata}}, we present both the simulated and experimentally measured coupling efficiency spectra of the frequency-splitting grating. The coupling efficiencies of the frequency-splitting grating computed using 2D FDTD are plotted in figure \textbf{\ref{fig:5_fsg_expdata}a}. At $1293~\mathrm{nm}$, the simulated coupling efficiencies into the left and right waveguides are $0.2937$ and $0.0008$ respectively, whereas at $1540~\mathrm{nm}$, the coupling efficiencies are $0.0027$ and $0.4544$. The measured signal intensity, normalized to source brightness, from the two output gratings is plotted in figure \textbf{\ref{fig:5_fsg_expdata}b}. The output grating coupling efficiencies were not measured, but due to the symmetry of the experimental setup, measurements of the two output ports are directly comparable. The experimental data from $1350~\mathrm{nm} - 1450~\mathrm{nm}$ is shaded to due to the presence of strong atmospheric water absorption lines \cite{ramcclatchey_dtic1972}. Fabry-perot fringing, arising from reflections between the frequency-splitting grating and output gratings, is visible in the measured spectra, with fringe spacing corresponding to the spacing between the gratings. The measured signals broadly match the simulated coupling efficiencies of the frequency-splitting grating.

The splitting ratio, defined as the ratio of power emitted from the two output ports, is plotted in figure \textbf{\ref{fig:5_fsg_expdata}c}. The fabricated device has a measured splitting ratio of $17\pm2~\mathrm{dB}$ at $1310~\mathrm{nm}$ and $12\pm2~\mathrm{dB}$ at $1540~\mathrm{nm}$, whereas the designed values at these wavelengths were $19.6~\mathrm{dB}$ and $22.2~\mathrm{dB}$ and respectively. The discrepancy between the simulated and measured splitting ratio is likely due to fabrication imperfections.

In conclusion, we have provided the first experimental demonstration of a nanophotonic device designed using our inverse design algorithm. In particular, we have implemented an efficient vertical-incidence wavelength-demultiplexing grating coupler, which cannot be designed by hand or by using parameter sweeps. By vastly opening up the parameter space for nanophotonic devices, inverse-design has broad implications for the future design of novel and compact nanophotonic components with full three-dimensional freedom.

\section*{Methods}

\subsection*{Optimization Algorithm and Electromagnetic Simulations}
Our inverse design algorithm uses a close analogue of steepest-descent optimization, and incorporates the MaxwellFDFD finite-difference frequency-domain solver to calculate local gradients \cite{wshin_jcp2012, wshin_maxwell_webpage}, as detailed in the Supplementary Information.

After completing the design process, the broadband performance of the device was calculated using 2D finite-difference time-domain (FDTD) simulations.

\subsection*{Fabrication}
The detailed grating dimensions are listed in Supplementary Table \textbf{1}. The devices were fabricated on Unibond$^{\mathrm{TM}}$ SmartCut$^{\mathrm{TM}}$ silicon-on-insulator (SOI) wafers obtained from SOITEC, with a nominal $220~\mathrm{nm}$ device layer and $3.0~\mathrm{\mu m}$ BOX layer. A JEOL JBX-6300FS electron beam lithography system was used to pattern $330~\mathrm{nm}$ of ZEP-52A electron beam resist spun on the samples. The pattern was then transferred to the Si device layer with a magnetically-enhanced reactive-ion etcher using a $\mathrm{H Br}/\mathrm{Cl_2}$ chemistry. Finally, the mask was stripped by sonicating in Microposit remover 1165.

\section*{Acknowledgements}
This work has been supported by the AFOSR MURI for Complex and Robust On-chip Nanophotonics (Dr. Gernot Pomrenke), grant number FA9550-09-1-0704. A.Y.P. also acknowledges support from the Stanford Graduate Fellowship. KGL acknowledges support from the Swiss National Science Foundation.

\section*{Author contributions}
A.Y.P. designed and fabricated the structures, simulated their performance, and performed the experiments. J.L. developed the inverse-design algorithm used in this paper. J.V. and T.B. provided theoretical and experimental guidance, and J.V. supervised the entire project.  K.G.L. and J.P. provided experimental support. All authors contributed to discussions.

\newpage
\begin{center}
\LARGE{Supplementary Information}
\end{center}

\section{Inverse Design Algorithm}
\subsection{Problem description}
Maxwell's equations in the frequency domain can be written as
\begin{align}
 \nabla \times \mu_0^{-1} \nabla \times \mathbf{E} - \omega^2 \epsilon \mathbf{E} = - i \omega \mathbf{J}
\end{align}
where $\mathbf{E}$ is the electric field, $\mathbf{J}$ is the current density, $\omega$ is the frequency, $\epsilon$ is the electric permittivity, and $\mu_0$ is the magnetic permeability of free space.

We specify the device performance by defining the mode conversion efficiency between a set of input modes and output modes. The input and output modes are specified by the user, and kept fixed during the optimization process. The input modes $i = 1 \ldots M$ are at frequencies $\omega_i$, and can be represented by equivalent current density distributions $\mathbf{J}_i$. We can then specify $N_i$ output modes of interest for each input mode $i$. The output mode electric fields $\mathcal{E}_{ij}$ are given over output surfaces $S_{ij}$, and the amplitude of each output mode should be bounded between $\alpha_{ij}$ and $\beta_{ij}$, where $j = 1 \ldots N_i$.

We are thus interested in finding $\epsilon$ and $\mathbf{E}_i$ which simultaneously satisfy
\begin{gather}
\nabla \times \mu_0^{-1} \nabla \times \mathbf{E}_i - \omega^2 \, \epsilon \, \mathbf{E}_i = - i \omega_i \mathbf{J}_i \\
\alpha_{ij} \leq \left| \iint_{S_{ij}} \mathcal{E}^\dagger_{ij} \cdot \mathbf{E}_i \mathrm{d}S \right| \leq \beta_{ij} 
\end{gather}
for $i = 1 \ldots M$ and $j = 1 \ldots N_i$. The permittivity $\epsilon$ is also subject to additional fabrication constraints.

This can be recast in the language of linear algebra by discretizing space and making the substitutions
\begin{alignat}{2}
\mathbf{E}_i & \rightarrow x_i &&\in \mathcal{C}^n \nonumber \\
\epsilon & \rightarrow z &&\in \mathcal{C}^n \nonumber \\
\nabla \times \mu_0^{-1} \nabla \times & \rightarrow D && \in \mathcal{C}^{n \times n} \nonumber \\
 - i \omega_i \mathbf{J}_i & \rightarrow b_i && \in \mathcal{C}^{n} \nonumber \\
\mathcal{E}_{ik} & \rightarrow c_{ij} &&\in \mathcal{C}^n.
\end{alignat}
This leaves us with the problem
\begin{gather}
D x_i - \omega_i^2 \diag(z) x_i - b_i = 0 \label{eqn:fsg_maxwell_linalg}\\
\alpha_{ij} \leq \left| c^\dagger_{ij} x_i \right| \leq \beta_{ij} 
\end{gather}
for $i = 1 \ldots M$ and $j = 1 \ldots N_i$. Here, $\diag\left(v\right)$ refers to the diagonal matrix whose diagonal entries are given by the vector $v$. For convenience, we further define
\begin{align}
A_i(z)   &\triangleq D - \omega_i^2 \diag(z) \nonumber \\
B_i(x_i) &\triangleq - \omega_i^2 \diag(x_i) \label{eqn:fsg_defAB}
\end{align}
which lets us rewrite equation (\ref{eqn:fsg_maxwell_linalg}) as
\begin{align}
0 = A_i(z) x_i - b_i = B_i(x_i) z + \left( D x_i + b_i \right).
\end{align}
The final problem we wish to solve is then
\begin{gather}
A_i(z) x_i - b_i = 0 \label{eqn:fsg_prob_maxwell}\\
\alpha_{ij} \leq \left| c^\dagger_{ij} x_i \right| \leq \beta_{ij}. \label{eqn:fsg_prob_constr}
\end{gather}

\subsection{Formulating the Optimization Problem}
We have previously developed two inverse design methods for designing linear optical devices: one which we call \textbf{objective-first}, and another which is an analogue of the \textbf{steepest-descent} strategy \cite{s_jlu_oe2013}. To design the WDM grating device demonstrated in this paper, we only used the steepest-descent based method, which is what we describe here.

The particular optimization problem we solve is
\begin{alignat}{3}
&\minimize  \quad && F\left(x_1, \ldots , x_M\right)  \nonumber \\
&\subjectto \quad && A_i(z) x_i - b_i = 0, \quad \mathrm{for} \; i = 1 \ldots N  \nonumber \\
&                  && z = m(p). 
\end{alignat}
Here, we constrain the fields to satisfy Maxwell's equations, parameterize the permittivity $z$ with $p \in \mathcal{R}^m$, and construct a penalty function
\begin{align}
F\left(x_1, \ldots , x_M\right) = \sum_{i = 1}^{M} f_i(x_i)
\end{align}
for violating our field constraints from equation (\ref{eqn:fsg_prob_constr}). The penalty $f_i(x_i)$ for each input mode is given by
\begin{align}
f_i = \sum_{j = 1}^{N_i} I_+\left( \left|  c_{ij}^\dagger x_i \right| - \alpha_{ij} \right) + I_+ \left(\beta_{ij} - \left| c_{ij}^\dagger x_i \right| \right)
\end{align}
where $I_+\left(u\right)$ is a relaxed indicator function,
\begin{align}
I_+\left(u\right) =
\begin{cases}
0, & u \geq 0 \\
\dfrac{1}{a} \left| u \right|^q, & \mathrm{otherwise}.
\end{cases}
\end{align}
Typically, we use $q = 2$ and $a = \max_i f_i(x_i) $.

\subsection{Optimization Algorithm}
We ensure that Maxwell's equations always satisfied, which implies that both the fields $x_1, \ldots, x_M$ and the field-constraint penalty function $F$ are a function of the permittivity $z$. On each iteration, we locally approximate our penalty function $F(z)$ with a quadratic function
\begin{align}
F(z) \approx Q(z) = \left\| P z - q \right\|^2
\end{align}
and solve the subproblem
\begin{alignat}{3}
&\minimize  \quad && Q(z)  \nonumber \\
&\subjectto \quad && z = m(p)
\end{alignat}
using steepest descent optimization. The structure parameter $p$ is sufficiently small that we can calculate the gradient $\nabla_p Q\left(m(p)\right)$ using brute force. The main computational cost of the algorithm lies in computing $Q(z)$. 

\subsection{Choice of $Q(z)$}
We choose a quadratic function $Q(z)$ of the form
\begin{align}
Q(z) &= \frac{1}{2} \left\| z - z_0 \right\|^2 + \kappa \nabla_z F^\dagger(z_0) \cdot \left( z - z_0 \right)  \nonumber \\
&= \frac{1}{2} \left\| z - \left( z_0 - \kappa \nabla_z F(z_0) \right) \right\|^2 + \left( \mathrm{const.} \right)
\end{align}
where $z_0$ is the value of $z$ from the previous iteration in the optimization process, and $\kappa \in \mathcal{R}$. The analogy with the steepest descent strategy is clear if we consider the minimum of $Q(z)$,
\begin{align}
\arg \min Q(z) = z_0 - \kappa \nabla_z F(z_0)
\end{align}
which is simply the steepest descent step with size $\kappa$.

We now consider how to compute the gradient $\nabla_z F$. Taking the derivative of equation (\ref{eqn:fsg_maxwell_linalg}), the discretized Maxwell's equations, with respect to $z$, we obtain
\begin{align}
D \frac{dx_i}{dz} - \omega_i^2 \diag(x_i) - \omega_i^2 \diag(z) \frac{dx_i}{dz} &= 0  \nonumber \\
\left( D - \omega_i^2 \diag(z) \right) \frac{dx_i}{dz} &= \omega_i^2 \diag (x_i)  \nonumber \\
A_i(z) \frac{dx_i}{dz} &= - B_i(x_i)
\end{align}
where we have used our definitions of $A_i$ and $B_i$ from (\ref{eqn:fsg_defAB}). The derivative of $x_i$ with respect to $z$ is then given by
\begin{align}
\frac{dx_i}{dz} &= - A^{-1}_i(z) B_i(x).
\end{align}
The gradient of the structure objective is thus
\begin{align}
\nabla_z F = \sum^M_{i = 1} \frac{d}{dz} f_i (x_i)
\end{align}
where
\begin{align}
\frac{d}{dz} f_i(x_i) = \frac{\partial f_i}{\partial x_i} \frac{d x_i}{dz} = -\frac{\partial f_i}{\partial x_i} A_i^{-1}(z) B_i(x_i) = - \left( A_i^{-\dagger}(z) \frac{\partial f_i^\dagger}{\partial x_i} \right)^\dagger B_i(x_i). \label{eqn:fsg_dfi_dxi}
\end{align}
Since $A_i$ and $B_i$ are large $n \times n$ matrices, we have rearranged the expression in the final step to require only a single matrix solve rather than $n$ solves.
The partial derivatives $\partial f_i / \partial x_i$ are given by
\begin{align}
\frac{\partial f_i}{\partial x_i} = \sum_{j = 1}^{N_i} \frac{\partial}{\partial x_i} I_+ \left( \left| c_{ij}^\dagger x_i \right| - \alpha_{ij} \right) + \frac{\partial}{\partial x_i} I_+ \left( \beta_{ij} - \left| c_{ij}^\dagger x_i \right| \right)
\end{align}
where 
\begin{align}
\frac{\partial}{\partial x_i} I_+ \left( \left| c_{ij}^\dagger x_i \right| - \alpha_{ij} \right)  &=  \frac{1}{2} \frac{\left(c_{ij}^\dagger x_i\right)^\ast}{\left| c_{ij}^\dagger x_i \right|}\, c_{ij}^\dagger \; \cdot
\begin{cases}
0, &  | c_{ij}^\dagger x_i | - \alpha_{ij} \geq 0 \\
\dfrac{q}{a} \left| | c_{ij}^\dagger x_i | - \alpha_{ij} \right|^{q - 1}, & \mathrm{otherwise}
\end{cases} \label{eqn:fsg_dIdx_1}\\
\frac{\partial}{\partial x_i} I_+ \left( \beta_{ij} - \left| c_{ij}^\dagger x_i \right| \right)  &=  \frac{1}{2} \frac{\left(c_{ij}^\dagger x_i\right)^\ast}{\left| c_{ij}^\dagger x_i \right|}\, c_{ij}^\dagger \; \cdot
\begin{cases}
0, &  \beta_{ij} - | c_{ij}^\dagger x_i | \geq 0 \\
\dfrac{q}{a} \left| \beta_{ij} - | c_{ij}^\dagger x_i | \right|^{q - 1}, & \mathrm{otherwise}.
\end{cases} \label{eqn:fsg_dIdx_2}
\end{align}
The absolute value function $\left|u\right|$ for $u \in \mathcal{C}$ is not analytic so the complex derivative does not exist. Instead, we have used the Wirtinger derivative \cite{s_rcgunning_1965, s_kbpetersen_matrixcook2012} of the absolute value function in (\ref{eqn:fsg_dIdx_1}) - (\ref{eqn:fsg_dIdx_2}), which is
\begin{align}
\frac{\partial}{\partial u} \left|u\right| = \frac{1}{2} \left(\frac{\partial}{\partial x} - i \frac{\partial}{\partial y} \right) \left| u \right| = \frac{u^\ast}{2 \left| u \right| }
\end{align}
where we have defined $u = x + i y$ for $x , y \in \mathcal{R}$.

\subsection{Parameterizing the structure}
The particular structure we designed consisted of two materials in the design area, with permittivities $\epsilon_1$ and $\epsilon_2$. We initially used a linear parameterization of $z = m(p)$,
\begin{align}
m(p) = z_{fixed} + \left(\epsilon_2 - \epsilon_1\right) S p
\end{align}
where $S \in \mathcal{R}^{n \times m}$, $0 \leq S_{kl} \leq 1$, and $0 \leq p_k \leq 1$. The portions of the structure which were kept fixed during the optimization process were described by $z_{fixed} \in \mathcal{C}$.

In the second step of the optimization, we converted to a level set representation of the structure using thresholding. When constructing $z$, we took care to apply anti-aliasing to the borders of the structure.

\subsection{Solving Maxwell's equations using FDFD}
We must efficiently solve Maxwell's equations in the frequency domain at the following points in our algorithm:
\begin{enumerate}
\item Evaluating the fields $x_i$ at the beginning of each iteration.
\item Solving the adjoint problem in equation (\ref{eqn:fsg_dfi_dxi}) to compute the gradient $\nabla_z F$ of the field-constraint penalty.
\end{enumerate}
This was done by using the MaxwellFDFD package for MATLAB \cite{s_wshin_jcp2012, s_wshin_maxwell_webpage}.

\newpage
\section{WDM grating dimensions}
\begin{table}[h]
\center
\textsf{
\begin{tabular}{r {r}@{$.$}{l} {r}@{$.$}{l} }
\toprule
$\mathbf{n}$ & \multicolumn{2}{c}{$t_n~(\mathrm{nm})$} & \multicolumn{2}{c}{$w_n~(\mathrm{nm})$} \\ 
\midrule
\textbf{1}   &   81&5   &   235&4 \\ 
\textbf{2}   &   83&4   &   235&5 \\ 
\textbf{3}   &   73&5   &   243&5 \\ 
\textbf{4}   &   81&9   &   218&8 \\ 
\textbf{5}   &   61&1   &   928&3 \\ 
\textbf{6}   &   147&1   &   85&2 \\ 
\textbf{7}   &   78&7   &   331&8 \\ 
\textbf{8}   &   164&4   &   450&8 \\ 
\textbf{9}   &   204&6   &   603&2 \\ 
\textbf{10}   &   192&6   &   257&0 \\ 
\textbf{11}   &   68&2   &   139&0 \\ 
\textbf{12}   &   130&6   &   399&5 \\ 
\textbf{13}   &   139&1   &   362&8 \\ 
\textbf{14}   &   98&5   &   156&8 \\ 
\textbf{15}   &   105&2   &   163&0 \\ 
\textbf{16}   &   75&0   &   665&7 \\ 
\textbf{17}   &   64&4   & \multicolumn{2}{c}{---} \\ 
\bottomrule
\end{tabular}
}

\caption{List of parameters for the grating coupler\label{tab:1_fsg_params}. The trench widths $\left(t_n\right)$ and spacings $\left(w_n\right)$ are indicated in figure \textbf{2a} of the main manuscript.}
\end{table}


\begin{thebibliography}{10}
\expandafter\ifx\csname url\endcsname\relax
  \def\url#1{\texttt{#1}}\fi
\expandafter\ifx\csname urlprefix\endcsname\relax\def\urlprefix{URL }\fi
\providecommand{\bibinfo}[2]{#2}
\providecommand{\eprint}[2][]{\url{#2}}

\bibitem{jsun_nat2013}
\bibinfo{author}{Sun, J.}, \bibinfo{author}{Timurdogan, E.},
  \bibinfo{author}{Yaacobi, A.}, \bibinfo{author}{Hosseini, E.~S.} \&
  \bibinfo{author}{R.Watts, M.}
\newblock \bibinfo{title}{Large-scale nanophotonic phased array}.
\newblock \emph{\bibinfo{journal}{Nature}} \textbf{\bibinfo{volume}{493}},
  \bibinfo{pages}{195 -- 199} (\bibinfo{year}{2013}).

\bibitem{pdtrinh_el1995}
\bibinfo{author}{Trinh, P.}, \bibinfo{author}{Yegnanarayanan, S.} \&
  \bibinfo{author}{Jalali, B.}
\newblock \bibinfo{title}{Integrated optical directional couplers in
  silicon-on-insulator}.
\newblock \emph{\bibinfo{journal}{Elect. Lett.}} \textbf{\bibinfo{volume}{31}},
  \bibinfo{pages}{2097 -- 2098} (\bibinfo{year}{1995}).

\bibitem{lbsoldano_jlt1995}
\bibinfo{author}{Soldano, L.~B.} \& \bibinfo{author}{Pennings, E. C.~M.}
\newblock \bibinfo{title}{Optical multi-mode interference devices based on
  self-imaging: principles and applications}.
\newblock \emph{\bibinfo{journal}{J. Lightw. Technol.}}
  \textbf{\bibinfo{volume}{13}}, \bibinfo{pages}{615 -- 627}
  (\bibinfo{year}{1995}).

\bibitem{temurphy_jlt2001}
\bibinfo{author}{Murphy, T.~E.}, \bibinfo{author}{Hastings, J.~T.} \&
  \bibinfo{author}{Smith, H.~I.}
\newblock \bibinfo{title}{Fabrication and characterization of narrow-band
  bragg-reflection filters in silicon-on-insulator ridge waveguides}.
\newblock \emph{\bibinfo{journal}{J. Lightw. Technol.}}
  \textbf{\bibinfo{volume}{19}}, \bibinfo{pages}{1938 -- 1942}
  (\bibinfo{year}{2001}).

\bibitem{pdumon_iptl2004}
\bibinfo{author}{Dumon, P.} \emph{et~al.}
\newblock \bibinfo{title}{Low-loss {SOI} photonic wires and ring resonators
  fabricated with deep {UV} lithography}.
\newblock \emph{\bibinfo{journal}{IEEE Photon. Technol. Lett.}}
  \textbf{\bibinfo{volume}{16}}, \bibinfo{pages}{1328 -- 1330}
  (\bibinfo{year}{2004}).

\bibitem{yshani_iqe1991}
\bibinfo{author}{Shani, Y.}, \bibinfo{author}{Henry, C.~H.},
  \bibinfo{author}{Kistler, R.~C.}, \bibinfo{author}{Kazarinov, R.~F.} \&
  \bibinfo{author}{Orlowsky, K.~J.}
\newblock \bibinfo{title}{Integrated optic adiabatic devices on silicon}.
\newblock \emph{\bibinfo{journal}{IEEE J. Quantum Electron.}}
  \textbf{\bibinfo{volume}{27}}, \bibinfo{pages}{556 -- 566}
  (\bibinfo{year}{1991}).

\bibitem{dtaillaert_jjap2006}
\bibinfo{author}{Taillaert, D.} \emph{et~al.}
\newblock \bibinfo{title}{Grating couplers for coupling between optical fibers
  and nanophotonic waveguides}.
\newblock \emph{\bibinfo{journal}{Jpn. J. Appl. Phys.}}
  \textbf{\bibinfo{volume}{45}}, \bibinfo{pages}{6071 -- 6077}
  (\bibinfo{year}{2006}).

\bibitem{agondarenko_oe2008}
\bibinfo{author}{Gondarenko, A.} \& \bibinfo{author}{Lipson, M.}
\newblock \bibinfo{title}{Low modal volume dipole-like dielectric slab
  resonator}.
\newblock \emph{\bibinfo{journal}{Opt. Express}} \textbf{\bibinfo{volume}{16}},
  \bibinfo{pages}{17689 -- 17694} (\bibinfo{year}{2008}).

\bibitem{ahakansson_oe2005}
\bibinfo{author}{H\r{a}kansson, A.} \& \bibinfo{author}{S\'{a}nchez-Dehesa, J.}
\newblock \bibinfo{title}{Inverse designed photonic crystal de-multiplex
  waveguide coupler}.
\newblock \emph{\bibinfo{journal}{Opt. Express}} \textbf{\bibinfo{volume}{13}},
  \bibinfo{pages}{5440 -- 5449} (\bibinfo{year}{2005}).

\bibitem{mminkov_sr2014}
\bibinfo{author}{Minkov, M.} \& \bibinfo{author}{Savona, V.}
\newblock \bibinfo{title}{Automated optimization of photonic crystal slab
  cavities}.
\newblock \emph{\bibinfo{journal}{Scientific Reports}}
  \textbf{\bibinfo{volume}{4}}, \bibinfo{pages}{5124} (\bibinfo{year}{2014}).

\bibitem{yma_oe2013}
\bibinfo{author}{Ma, Y.} \emph{et~al.}
\newblock \bibinfo{title}{Ultralow loss single layer submicron silicon
  waveguide crossing for {SOI} optical interconnect}.
\newblock \emph{\bibinfo{journal}{Opt. Express}} \textbf{\bibinfo{volume}{21}},
  \bibinfo{pages}{29374 -- 29382} (\bibinfo{year}{2013}).

\bibitem{sboyd_2004}
\bibinfo{author}{Boyd, S.} \& \bibinfo{author}{Vandenberghe, L.}
\newblock \emph{\bibinfo{title}{{C}onvex {O}ptimization}}
  (\bibinfo{publisher}{Cambridge University Press}, \bibinfo{year}{2004}).

\bibitem{jsjensen_apl2004}
\bibinfo{author}{Jensen, J.~S.} \& \bibinfo{author}{Sigmund, O.}
\newblock \bibinfo{title}{Systematic design of photonic crystal structures
  using topology optimization: Lowloss waveguide bends}.
\newblock \emph{\bibinfo{journal}{Appl. Phys. Lett.}}
  \textbf{\bibinfo{volume}{84}}, \bibinfo{pages}{2022} (\bibinfo{year}{2004}).

\bibitem{piborel_oe2004}
\bibinfo{author}{Borel, P.~I.} \emph{et~al.}
\newblock \bibinfo{title}{Topology optimization and fabrication of photonic
  crystal structures}.
\newblock \emph{\bibinfo{journal}{Opt. Express}} \textbf{\bibinfo{volume}{12}},
  \bibinfo{pages}{1996 -- 2001} (\bibinfo{year}{2004}).

\bibitem{amutapcica_eo2009}
\bibinfo{author}{Mutapcica, A.}, \bibinfo{author}{Boyd, S.},
  \bibinfo{author}{Farjadpour, A.}, \bibinfo{author}{Johnson, S.~G.} \&
  \bibinfo{author}{Avnielb, Y.}
\newblock \bibinfo{title}{Robust design of slow-light tapers in periodic
  waveguides}.
\newblock \emph{\bibinfo{journal}{Eng. Optimiz.}}
  \textbf{\bibinfo{volume}{41}}, \bibinfo{pages}{365–384}
  (\bibinfo{year}{2009}).

\bibitem{jjensen_lpr2011}
\bibinfo{author}{Jensen, J.~S.} \& \bibinfo{author}{Sigmund, O.}
\newblock \bibinfo{title}{Topology optimization for nano-photonics}.
\newblock \emph{\bibinfo{journal}{Laser Photonics Rev.}}
  \textbf{\bibinfo{volume}{5}}, \bibinfo{pages}{308 -- 321}
  (\bibinfo{year}{2011}).

\bibitem{lalau-keraly_oe2013}
\bibinfo{author}{Lalau-Keraly, C.~M.}, \bibinfo{author}{Bhargava, S.},
  \bibinfo{author}{Miller, O.~D.} \& \bibinfo{author}{Yablonovitch, E.}
\newblock \bibinfo{title}{Adjoint shape optimization applied to electromagnetic
  design}.
\newblock \emph{\bibinfo{journal}{Opt. Express}} \textbf{\bibinfo{volume}{21}},
  \bibinfo{pages}{21693 -- 21701} (\bibinfo{year}{2013}).

\bibitem{aniederberger_oe2014}
\bibinfo{author}{Niederberger, A. C.~R.}, \bibinfo{author}{Fattal, D.~A.},
  \bibinfo{author}{Gauger, N.~R.}, \bibinfo{author}{Fan, S.} \&
  \bibinfo{author}{Beausoleil, R.~G.}
\newblock \bibinfo{title}{Sensitivity analysis and optimization of
  sub-wavelength optical gratings using adjoints}.
\newblock \emph{\bibinfo{journal}{Opt. Express}} \textbf{\bibinfo{volume}{22}},
  \bibinfo{pages}{12971 -- 12981} (\bibinfo{year}{2014}).

\bibitem{jlu_oe2013}
\bibinfo{author}{Lu, J.} \& \bibinfo{author}{Vu\v{c}kovi\'{c}, J.}
\newblock \bibinfo{title}{Nanophotonic computational design}.
\newblock \emph{\bibinfo{journal}{Opt. Express}} \textbf{\bibinfo{volume}{21}},
  \bibinfo{pages}{13351 -- 13367} (\bibinfo{year}{2013}).

\bibitem{dabmiller_oe2012}
\bibinfo{author}{Miller, D. A.~B.}
\newblock \bibinfo{title}{All linear optical devices are mode converters}.
\newblock \emph{\bibinfo{journal}{Opt. Express}} \textbf{\bibinfo{volume}{20}},
  \bibinfo{pages}{23985 -- 23993} (\bibinfo{year}{2012}).

\bibitem{gtreed_2008}
\bibinfo{author}{Reed, G.~T.}
\newblock \emph{\bibinfo{title}{Silicon Photonics: The State of the Art}}
  (\bibinfo{publisher}{John Wiley \& Sons}, \bibinfo{year}{2008}).

\bibitem{groelkens_oe2007}
\bibinfo{author}{Roelkens, G.}, \bibinfo{author}{Thourhout, D.~V.} \&
  \bibinfo{author}{Baets, R.}
\newblock \bibinfo{title}{Silicon-on-insulator ultra-compact duplexer based on
  a diffractive grating structure}.
\newblock \emph{\bibinfo{journal}{Opt. Express}} \textbf{\bibinfo{volume}{15}},
  \bibinfo{pages}{10091 -- 10096} (\bibinfo{year}{2007}).

\bibitem{wshin_jcp2012}
\bibinfo{author}{Shin, W.} \& \bibinfo{author}{Fan, S.}
\newblock \bibinfo{title}{Choice of the perfectly matched layer boundary
  condition for frequency-domain {M}axwell's equations solvers}.
\newblock \emph{\bibinfo{journal}{J. Comput. Phys.}}
  \textbf{\bibinfo{volume}{231}}, \bibinfo{pages}{3406–3431}
  (\bibinfo{year}{2012}).

\bibitem{wshin_maxwell_webpage}
\bibinfo{author}{Shin, W.}
\newblock \bibinfo{title}{{M}axwell{FDFD} {W}ebpage} (\bibinfo{year}{2014}).
\newblock \urlprefix\url{https://github.com/wsshin/maxwellfdfd}.

\bibitem{sosher_2003}
\bibinfo{author}{Osher, S.} \& \bibinfo{author}{Fedkiw, R.}
\newblock \emph{\bibinfo{title}{Level Set Methods and Dynamic Implicit
  Surfaces}} (\bibinfo{publisher}{Springer}, \bibinfo{year}{2003}).

\bibitem{ckopp_ijqe2011}
\bibinfo{author}{Kopp, C.} \emph{et~al.}
\newblock \bibinfo{title}{Silicon photonic circuits: {O}n-{CMOS} integration,
  fiber optical coupling, and packaging}.
\newblock \emph{\bibinfo{journal}{IEEE J. Sel. Topics Quantum Electron.}}
  \textbf{\bibinfo{volume}{17}}, \bibinfo{pages}{498 -- 509}
  (\bibinfo{year}{2011}).

\bibitem{ramcclatchey_dtic1972}
\bibinfo{author}{McClatchey, R.~A.}, \bibinfo{author}{Fenn, R.~W.},
  \bibinfo{author}{Selby, J. E.~A.}, \bibinfo{author}{Volz, F.~E.} \&
  \bibinfo{author}{Garing, J.~S.}
\newblock \bibinfo{title}{Optical properties of the atmosphere (third
  edition)}.
\newblock \bibinfo{type}{Scientific} \bibinfo{number}{AFCRL-72-0497},
  \bibinfo{institution}{Air Force Cambridge Research Laboratories, Hanscom AFB,
  United States Air Force}, \bibinfo{address}{L. G. Hanscom Field Bedford,
  Massachusetts 01730} (\bibinfo{year}{1972}).

\end{thebibliography}

\begin{thebibliography}{1}
\expandafter\ifx\csname url\endcsname\relax
  \def\url#1{\texttt{#1}}\fi
\expandafter\ifx\csname urlprefix\endcsname\relax\def\urlprefix{URL }\fi
\providecommand{\bibinfo}[2]{#2}
\providecommand{\eprint}[2][]{\url{#2}}

\bibitem{s_jlu_oe2013}
\bibinfo{author}{Lu, J.} \& \bibinfo{author}{Vu\v{c}kovi\'{c}, J.}
\newblock \bibinfo{title}{Nanophotonic computational design}.
\newblock \emph{\bibinfo{journal}{Opt. Express}} \textbf{\bibinfo{volume}{21}},
  \bibinfo{pages}{13351 -- 13367} (\bibinfo{year}{2013}).

\bibitem{s_rcgunning_1965}
\bibinfo{author}{Gunning, R.~C.} \& \bibinfo{author}{Rossi, H.}
\newblock \emph{\bibinfo{title}{Analytic functions of several complex
  variables}}.
\newblock Prentice-Hall series in modern analysis
  (\bibinfo{publisher}{Prentice-Halls}, \bibinfo{year}{1965}).

\bibitem{s_kbpetersen_matrixcook2012}
\bibinfo{author}{Petersen, K.~B.} \& \bibinfo{author}{Pedersen, M.~S.}
\newblock \emph{\bibinfo{title}{The {M}atrix {C}ookbook}}.
\newblock \bibinfo{organization}{Technical University of Denmark}
  (\bibinfo{year}{2012}).

\bibitem{s_wshin_jcp2012}
\bibinfo{author}{Shin, W.} \& \bibinfo{author}{Fan, S.}
\newblock \bibinfo{title}{Choice of the perfectly matched layer boundary
  condition for frequency-domain {M}axwell's equations solvers}.
\newblock \emph{\bibinfo{journal}{J. Comput. Phys.}}
  \textbf{\bibinfo{volume}{231}}, \bibinfo{pages}{3406–3431}
  (\bibinfo{year}{2012}).

\bibitem{s_wshin_maxwell_webpage}
\bibinfo{author}{Shin, W.}
\newblock \bibinfo{title}{{M}axwell{FDFD} {W}ebpage} (\bibinfo{year}{2014}).
\newblock \urlprefix\url{https://github.com/wsshin/maxwellfdfd}.

\end{thebibliography}
\end{document}